\pdfoutput=1

\documentclass[11pt,twoside,a4paper,cmspaper,final,collab]{cms-tdr}
\def\svnVersion{167114}\def\cmsCernNo{2012-159}\def\cmsCernDate{\today}\def\cmsMessage{Submitted to Physical Review Letters}

\begin{document}\cmsNoteHeader{SUS-11-010}

\hyphenation{had-ron-i-za-tion}
\hyphenation{cal-or-i-me-ter}
\hyphenation{de-vices}

\RCS$Revision: 134593 $
\RCS$HeadURL: svn+ssh://svn.cern.ch/reps/tdr2/papers/SUS-11-010/trunk/SUS-11-010.tex $
\RCS$Id: SUS-11-010.tex 134593 2012-07-03 16:04:35Z alverson $

\newcommand{\intLumi}{\ensuremath{4.98\fbinv} }
\providecommand{\met}{\ETm}
\newcommand{\pp}{\Pp{}\Pp}
\newlength\cmsFigWidth
\ifthenelse{\boolean{cms@external}}{\setlength\cmsFigWidth{0.95\columnwidth}}{\setlength\cmsFigWidth{0.7\textwidth}}

\cmsNoteHeader{SUS-11-010} 
\title{Search for new physics with same-sign isolated dilepton events with jets and missing transverse energy}

\date{\today}

\abstract{
A  search for new physics is performed in events with two same-sign
isolated leptons, hadronic jets,
and missing transverse energy in the final state.
The analysis is based on a data sample corresponding to an integrated luminosity of \intLumi produced
in $\pp$ collisions at a center-of-mass energy
of 7\TeV collected by the CMS experiment at the LHC. This constitutes a factor of 140 increase in
integrated luminosity over previously published results.
The observed yields agree with the standard model
predictions and thus no evidence for new physics is found.
The observations are used to set upper limits on possible new physics
contributions and to constrain supersymmetric models.  To facilitate the interpretation of the data
in a broader range of new physics scenarios, information on the event selection, detector
response, and efficiencies is provided.
}

\hypersetup{%
pdfauthor={CMS Collaboration},%
pdftitle={Search for new physics with same-sign isolated dilepton events with jets and missing transverse energy},%
pdfsubject={CMS},%
pdfkeywords={CMS, physics, SUSY, analysis, 2011}}

\maketitle 

The standard model (SM) is a very successful theory of elementary
particles and their interactions.
It is generally believed that new physics (NP)  could
manifest itself  at the TeV scale.  Supersymmetry (SUSY)
is one of these attractive possibilities.   It leads to gauge
coupling unification at very high energy, provides a mechanism to mitigate large
radiative corrections to the Higgs mass
and, in its R-parity-conserving~\cite{Fayet:1976cr} realization, can provide a dark matter candidate.
A comprehensive program of searches
for the production of supersymmetric particles has been underway since 2010
at the Large Hadron Collider (LHC).  Since SUSY models vary widely, these searches
target
a broad range of possible final states,
including purely hadronic states~\cite{AtlasHad,CMSHad}, leptonic states
with one lepton~\cite{Atlas1lep,CMS1lep}, two leptons
of the opposite sign~\cite{AtlasOS,CMSOS}, two leptons of the same sign~\cite{AtlasOS,sus10004},
and three or more leptons~\cite{CMSmult}; as well as
photonic final states~\cite{Atlasgamma,CMSgamma}.

In this Letter we report on a search for NP based on isolated same-sign (SS) dileptons, missing transverse energy (\met), and hadronic jets. In SUSY SS
dileptons can arise, for example, from pair production of colored super-partners (gluinos and/or squarks), with a lepton in the decay chain of each primary
SUSY particle~\cite{Barnett:1993ea,Guchait:1994zk,Baer:1995va}; more generally, this signature is sensitive to final states with same-sign $\PW$ bosons and/or top
quarks~\cite{Cheng:2002ab,Jung:2010,Berger:2011, Contino:2008hi, Keung:1983uu, Han:2009}. The rarity of SS dileptons
in the SM makes a NP search in this final state particularly attractive.

All types of charged leptons, $\Pe$, $\Pgm$, and
hadronically decaying $\Pgt$s, are included in our search.
These final states are indicators of the possible presence of SUSY particles
as well as other possible NP
scenarios.
The results are based on a data sample corresponding to
$4.98 \pm 0.11\fbinv$ of $\pp$ collisions at a center-of-mass energy of 7\TeV
collected in 2011 by the Compact Muon Solenoid (CMS)~\cite{CMS:2008zzk} experiment at the LHC.
This study results in a major improvement in sensitivity with respect
to the search performed with data collected in 2010~\cite{sus10004} because of the 140-fold increase
in the integrated luminosity of the data sample.  These results are interpreted using
the constrained minimal supersymmetric extension of the standard model (CMSSM)~\cite{cmssm}.
In addition, this analysis provides information on the event selection
and detector response in order to
facilitate the application of our results to a broader
range of NP scenarios.

A detailed description of the CMS detector is found elsewhere~\cite{CMS:2008zzk}.
Its central feature is a superconducting solenoid providing an axial magnetic field of 3.8\unit{T}.
Muons are measured in gas detectors embedded in the steel return yoke of the magnet,
while all other particle detection systems are located inside the bore of the solenoid.
Charged particle trajectories are measured by a silicon pixel and strip tracker system,
covering
$|\eta|<2.5$,
where the pseudorapidity is defined as $\eta = - \ln[\tan{\theta/2}]$,
and $\theta$ is the polar angle with respect to the counterclockwise beam direction.
A crystal electromagnetic calorimeter (ECAL) and a brass/scintillator hadronic
calorimeter surround the tracker volume. In addition, the CMS detector has an extensive forward calorimeter and nearly hermetic $4\pi$
coverage. The CMS trigger consists of a two-stage system.
The first level of the CMS trigger system, composed of custom hardware processors, uses
information from the calorimeters and muon detectors to select a subset of the events. The
High Level Trigger processor farm further decreases the event rate from around 100\unit{kHz}
to around 300\unit{Hz}, before data storage.

All lepton candidates are required to
have $|\eta|<2.4$ and to be consistent with a common interaction vertex.
Muon candidates are reconstructed~\cite{MUOPAS} by matching
tracks in the silicon detector to signals in the muon system.
The reconstruction of muons is refined further
by imposing track quality and calorimeter energy deposition requirements.
Electron candidates are reconstructed~\cite{EGMPAS} starting from a
cluster of energy deposits in the
ECAL, which is then matched to signals in the silicon tracker.
The energy shower in the ECAL must have a shape consistent with expectations for
electron showers and its position is required to be well matched to
the extrapolated track.
Both electrons and muons are required to be isolated from other
activity in the event.
This is achieved using a scalar sum of transverse track momenta and transverse calorimeter energy deposits,
within $\Delta R \equiv\sqrt{{ (\Delta\phi)^2+(\Delta\eta)^2}} <0.3$ of the candidate's direction, where $\phi$
is the azimuthal angle. The sum is required to be less than 15\% of the candidate's transverse momentum ($\pt$).
Hadronic $\Pgt$ candidates are reconstructed using the Hadron plus Strip algorithm~\cite{PFT-10-XXX}.
We select isolated hadronic $\Pgt$ candidates with
one or three charged hadrons in a narrow cone around the $\Pgt$ direction.

Jets and \met~are reconstructed using the particle-flow technique~\cite{PFT-09-001,PFT-10-002}.
For jet clustering, the anti-$k_\mathrm{T}$ algorithm is used
with the distance parameter $R = 0.5$~\cite{anti-kt}.
We require selected jets to have $\pt >40\GeV$ and $|\eta|<2.5$ to be considered for analysis.
The $\HT$ is defined to be the scalar sum of the $\pt$ of all
selected jets whose angular distance to the nearest lepton satisfies ${ \Delta R > 0.4}$.
Events are required to have two same-sign leptons and at least two jets. A minimum dilepton invariant mass of 8\GeV is required
in order to suppress the low-mass dilepton background. Events having a third lepton are removed if two of the leptons form a $\cPZ$
boson candidate with an invariant mass within $\pm 15$\GeV of the $\cPZ$ boson mass.

Three selection strategies are followed to maximize the sensitivity
to the presence of NP.
The first one is to use a fully efficient dilepton and  $\HT$ based trigger
in the $\Pe\Pe$, $\Pgm\Pgm$, and $\Pe\Pgm$ channels with $\pt^\Pgm > 5\GeV$ and $\pt^\Pe >10\GeV$,
and a requirement of $\HT>200\GeV$ applied to the offline reconstructed objects.
The second strategy trades an increased lepton $\pt$ threshold against a reduced $\HT$ threshold.
Here both leptons are required to have $\pt > 10\GeV$ and at least one to have $\pt > 20\GeV$.
Such events are collected with a purely leptonic trigger with no requirement on $\HT$.
The third strategy focuses on $\Pgt\Pe, \Pgt \Pgm$, and $\Pgt \Pgt$ final states with $\pt^\Pgm >5\GeV$,
$\pt^\Pe >10\GeV$, and $\pt^\Pgt >  15\GeV$.
Triggers for hadronic  $\Pgt$-leptons typically lead to high rates.
For this reason dedicated triggers are used that rely on significant $\HT$ and \met,
in addition to the presence of a single lepton or two hadronic $\Pgt$-leptons.

\begin{figure}[hbtp]
\begin{center}
\includegraphics[width=0.48\textwidth]{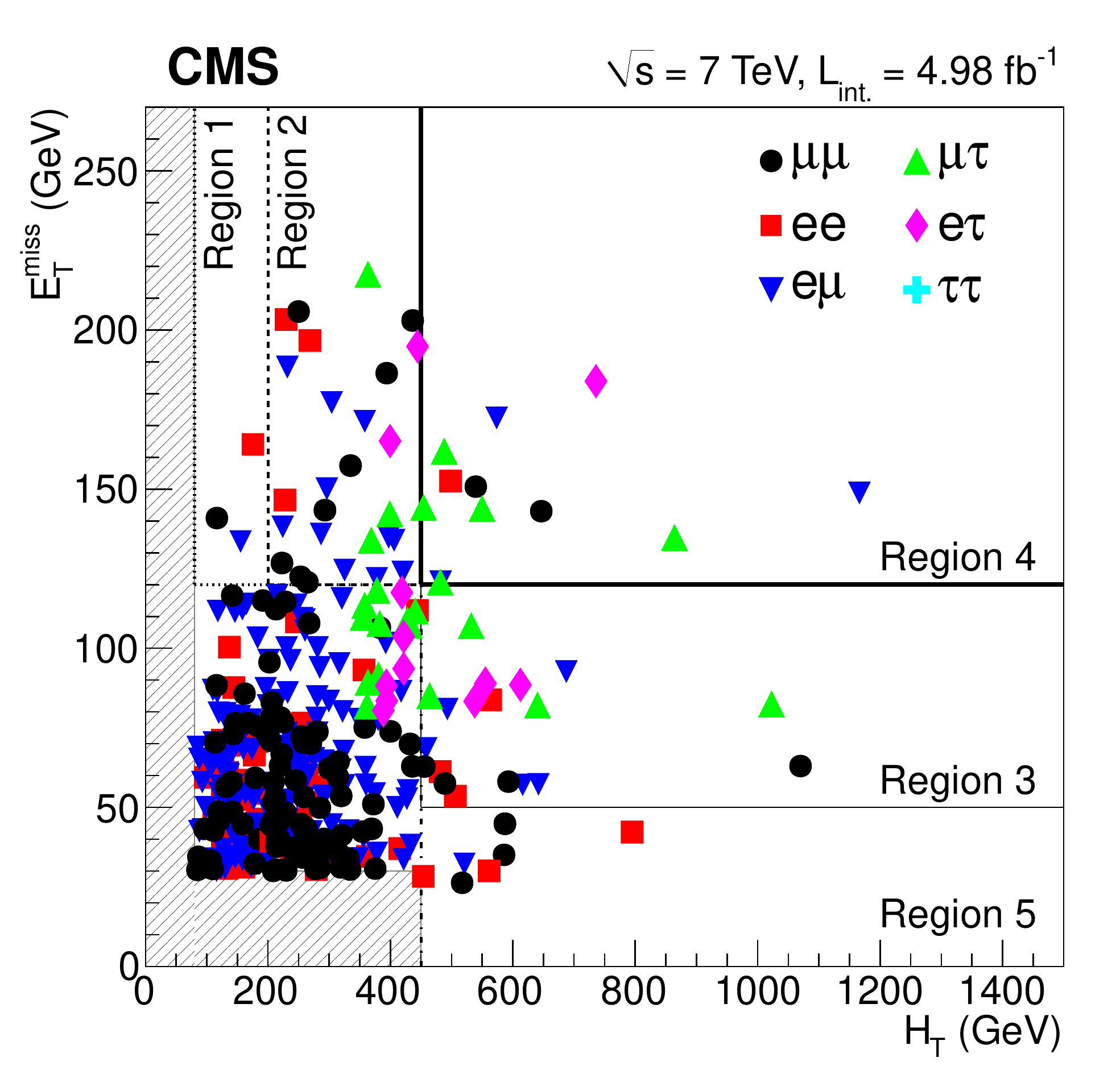}
\caption{\label{fig:scatter} Selected SS dilepton events in the various search regions displayed
in the \HT, \met~plane.}
\end{center}
\end{figure}

Using R-parity--conserving SUSY as a guiding example, we note that the simplest incarnation of the
topology probed by this analysis involves three distinct mass scales.  In this example,
these masses would belong to the gluino, chargino, and lightest SUSY particle (LSP).  The mass differences
of these particles can strongly influence the kinematics of the final state objects, hence
affecting several main observables used in this analysis: lepton $\pt$, $\HT$, and \met.
Therefore, in order to maximize the sensitivity of our analysis to a variety of NP scenarios, we
define multiple search regions in the ($\HT$, \met)  plane~:
Region~1 ($\HT>80\GeV$, \met$>120\GeV$), Region~2 ($\HT>200\GeV$, \met$>120 \GeV$), Region~3 ($\HT>450\GeV$, \met$>50$ \GeV),
Region~4 ($\HT>450\GeV$ and \met$>120\GeV$), and Region~5 ($\HT>450\GeV$ and \met$>0\GeV$).
The $\HT$ requirements of $200$ GeV and $450$ GeV
are also motivated in part by trigger thresholds.
A scatter plot of events observed in these search regions is shown in Fig.~\ref{fig:scatter}.

The background for the same-sign dilepton topology has three components:
irreducible background from rare SM processes;
leptons resulting from semileptonic decays within a jet,
or jets mimicking leptons in events with zero or one genuine isolated lepton;
and opposite-sign dilepton events where the charge of one of the two leptons has been mismeasured.

The irreducible backgrounds are dominated by $\cPqt\cPaqt+\Wpm\slash\cPZ$, $\Wpm\Wpm\cPq\cPq$, and $\Wpm\cPZ$ production,
combining in similar parts to about 95\% of the total. The remaining contributions originate
from processes such as tri-boson and $\cPZ\cPZ$ production, $\Wpm\cPZ+\cPgg$, and
double-parton scattering $2\times(\cPq\cPaq^\prime\rightarrow \Wpm)$, in decending order of
importance.  All irreducible backgrounds are estimated using leading-order Monte Carlo simulation
normalized to the next-to-leading order (NLO) production cross sections.
Events are generated with the \MADGRAPH~\cite{MadGraph} event generator and then passed on to \PYTHIA~\cite{PYTHIA} for parton shower
and hadronization. The generated events are processed by the CMS event simulation and the
same chain of reconstruction programs used for collision data.
A 50\% systematic uncertainty is assigned to this irreducible background prediction.
These processes constitute 35-75\% of the total background, depending on the search region.

The background due to lepton candidates originating from jets, hereafter referred as non-prompt,
forms 20-60\% of the total background.
Such candidates can be genuine leptons, for example from heavy-flavor decays,
hadrons reconstructed as leptons, or jets fluctuating to give hadronic $\Pgt$ signatures.
We have developed and validated a set of techniques to measure this background from data.
In each case, a tag-and-probe method is applied to a control sample rich in two-jet events
containing leptons selected with loose
 requirements to measure the conditional probability that the probe jet
 yields a candidate passing tight lepton requirements.  This probability, measured as a function of
jet kinematics and event characteristics, is then applied to signal sidebands to estimate non-prompt lepton
backgrounds. This suite of techniques encompasses a range of control samples, jet tags, lepton
requirements, and variations in the jet kinematics to provide independent and complementary
assessments of 50\% systematic uncertainties.  Full details are given in Ref.~\cite{sus10004}.
At least two techniques are used in all non-$\Pgt$ dilepton modes and they yield consistent
results within their respective uncertainties.

We quantify backgrounds from events with lepton charge mis-reconstruction by analyzing SS
$\Pe\Pe$ or $\Pgt\Pgt$ events inside the $\cPZ$ mass peak~\cite{sus10004}.
This background forms less than 5\% of the total background across all search regions. The charge mis-reconstruction
probability for muons is of the order of $10^{-5}$ and can be neglected.

We determine the performance of the background prediction methods using the low $\HT$ and low \met~region in the data that
is expected to be dominated by SM events. We find good agreement between observed yields and the predicted background.

\begin{figure}[hbtp]
\begin {center}
\includegraphics[width=0.48\textwidth]{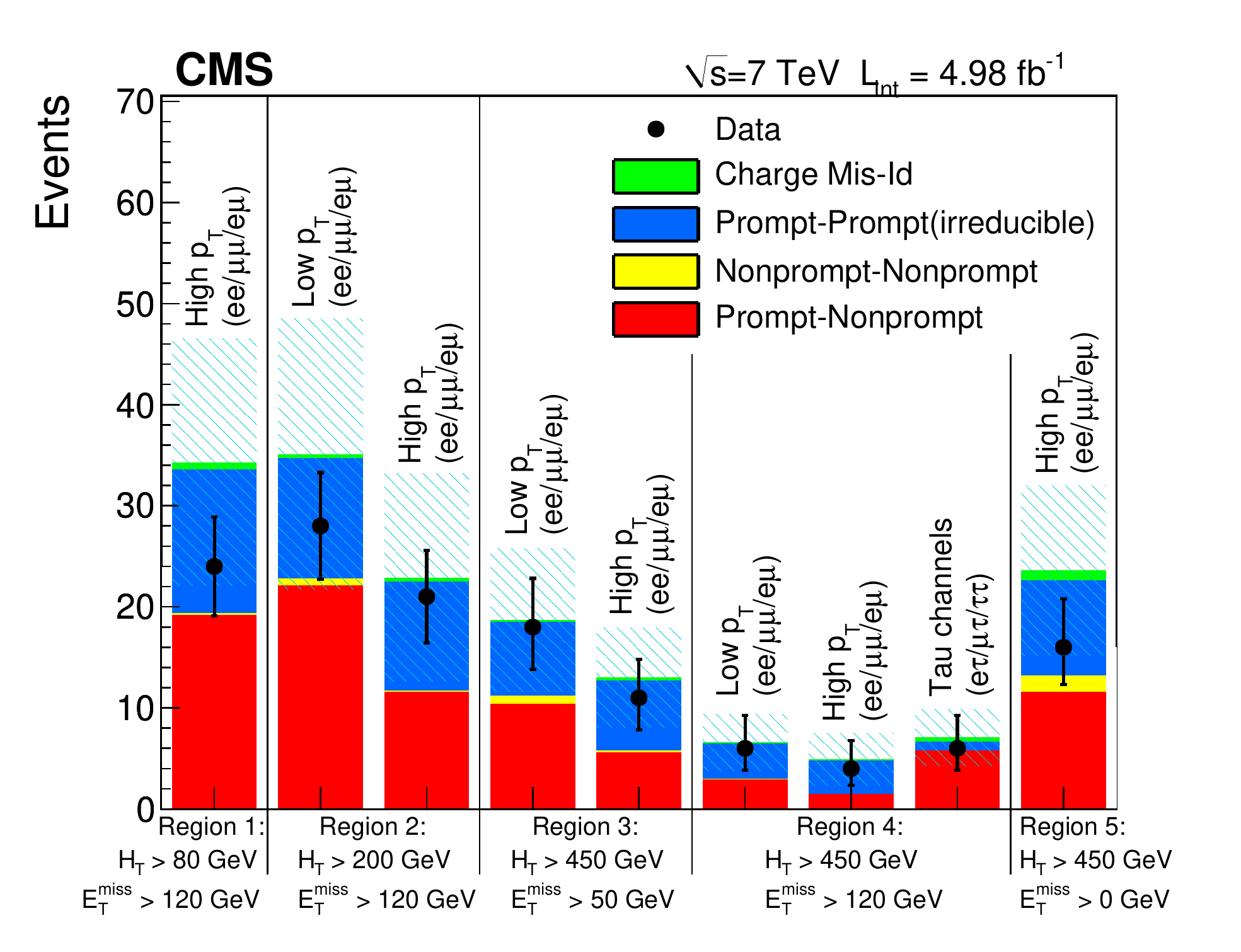}
\caption{\label{fig:predictions} Summary of background predictions and observed yields in the various search regions. Leptons
from decays of $\PW$, $\cPZ$, and NP particles are referred to as prompt leptons. The hatched bands represent the
total uncertainty on the background predictions.}
\end{center}
\end{figure}

We show the predicted background contributions from each source mentioned above as well as the observed event yields in Fig.~\ref{fig:predictions}
and summarize them in Table~\ref{tab:results} for each search region.  The beam related multiple interactions do not alter these results.
There is no evidence of an excess over the expected SM predictions.
This measurement is used together with the uncertainty on the signal acceptance to set an upper limit (UL) on the contribution from NP events.

\newlength\cmsStrutHeight
\begin{table}[hbtp]
\topcaption{\label{tab:results} Observed number of events in data compared to the predicted background
yields for the considered search regions.
The uncertainties include the statistical and systematic components
added in quadrature with correlations taken into account.
The 95\% CL upper limit (UL) on the contribution from NP events is also given.}
\begin{center}\footnotesize
\setlength{\extrarowheight}{2pt}
\setlength\cmsStrutHeight{3ex plus 9pt}
\begin{tabular}{cccccc}\hline \hline
Region  & \multicolumn{3}{c|}{Mode or $\pt$ threshold}     & Total  & UL
\\ \hline \hline
\rule{0pt}{\cmsStrutHeight} &\multicolumn{3}{c|}{ High $\pt: \pt^{\ell1, \ell2} > 20, 10$\GeV } & & \\ \cline{2-4}
        & $\Pe\Pe$          & $\Pgm\Pgm$      & $\Pe\Pgm$        &               & \\ \hline
1       &{$6.8 \pm 2.7$}        &{$8.6 \pm 3.3$}        &{$18.7 \pm 6.9$ }      &{$34.1 \pm 12.2$}  \\
        &  {5}          & {7}           & {12}          & {24}                  & {13.7} \\ \hline
2       &{$4.3 \pm1.9$}         &{$6.1\pm2.4$}  &{$12.2 \pm 4.6$}       &{$22.6\pm8.3$}  \\
        & {4}           & {6}           & {11}          & {21}                  & {15.1} \\ \hline
3       &{$3.8 \pm1.7$}         &{$3.1\pm1.4$}  &{$6.1\pm2.4$}  &{$13.0\pm4.9$}  \\
        & {4}           & {2}           & {5}           & {11}                  & {9.6} \\ \hline
4       &{$1.1 \pm1.1$}  &{$1.2 \pm1.2$}  &{$2.6 \pm1.4$}  &{$4.9\pm2.6$}   \\
        & {1}            & {0}            & {3}           & {4}           & {6.2} \\ \hline
5       &{$9.1 \pm3.6$} &{$4.7 \pm1.9$}  &{$9.8 \pm3.7$}  &{$23.6 \pm8.4$} \\
        & {7}           & {4}            & {5}            & {16}          & {10.4} \\ \hline \hline
       & \multicolumn{3} {c|} { Low $\pt: \pt^{\Pe, \Pgm} > 10, 5$\GeV }     & & \\ \cline{2-4}
        & $\Pe\Pe$          & $\Pgm\Pgm$      & $\Pe\Pgm$        &               & \\ \hline
2       & {$4.4 \pm 1.8$} &{$14.1 \pm6.0 $}     &{$16.5 \pm6.4 $}       &{$35.0 \pm 13.4$}  \\
        & {4}           & {10}          & {14}          & {28}                  & {16.9} \\ \hline
3       &{$3.4\pm 1.6$}         &{$6.5\pm2.8$}  &{$8.9\pm3.6$}  &{$18.8\pm 7.1$}  \\
        &  {4}          & {6}           & {8}           & {18}                  & {14.0} \\ \hline
4       &{$1.0 \pm 0.8$}        &{$2.4\pm1.2$}  &{$3.2\pm1.5$}  &{$6.6\pm2.8$}  \\
        &  {1}          & {2}           & {3}           & {6}           & {7.4} \\ \hline \hline

        & \multicolumn{3} {c|}  { Tau channels : $\pt^{\Pe, \Pgm, \Pgt} > 10, 5, 15$\GeV } & & \\ \cline{2-4}
        & $\Pe\Pgt$       & $\Pgm\Pgt$     & $\Pgt\Pgt$    & & \\ \hline
4       &{$2.6\pm1.0$}  &{$4.4\pm2.2$}  &{$0.0\pm0.1$}  &{$7.1\pm2.8$}  \\
        &  {1}          & {5}           & {0}           & {6}           & {7.1} \\ \hline
\hline
\end{tabular}
\end{center}
\end{table}

We measure the electron and muon selection efficiencies in data
and simulation using $\cPZ$ events
to derive simulation-to-data correction factors.
The uncertainty on the combined lepton selection efficiency
decreases with lepton $\pt$, from 5\% at the lowest $\pt$ to 3\% above $20$\gev.
We assign an additional 5\% systematic uncertainty per lepton to cover potential mismodeling of
the lepton isolation efficiency due to varying hadronic activity in signal events.
We estimate in a sample of $\cPZ\rightarrow \Pgt\Pgt$ events the uncertainty on the $\Pgt$ selection and reconstruction efficiency to be 10\%~\cite{PFT-10-XXX}.

We conservatively choose to attribute a flat uncertainty of 7.5\%  to the energy measurement
of all jets as well as to the hadronic component used for the \met~observable.
The cumulative effect of this uncertainty on the signal acceptance is
intrinsically model dependent. We observe uncertainties below 3\% for models with characteristically
high $\HT$ scales, well above the $\HT$ requirements.
For models with characteristic $\HT$ scales
near or below the $\HT$ requirements, uncertainties due to jet energy calibration can be as high as 30\%.

The theoretical uncertainties on the signal acceptance due to the modeling of
initial- and final-state radiation and knowledge of
the parton distribution functions are estimated to be 2\%.
Using the LM6 benchmark model (CMSSM point with $m_0= 85$\gev, $m_{1/2}=400$\gev,
$\tan\beta=10$, $A_0=0$, and $\mu>0$)~\cite{PTDR2} as a signal model, the total experimental and
theoretical uncertainties in the signal yield add up to 14\% or 20\% depending on the
search region. This includes a 2.2\%  systematic uncertainty in the integrated luminosity~\cite{lumi}.

We  set a 95\% confidence level (CL) upper limits on the number of observed events using
the modified frequentist construction CL$_{s}$ method~\cite{Read:2002hq,Junk:1999kv,ATL-PHYS-PUB-2011-011}.
We assume log-normal distributions for the efficiency and background uncertainties.
As a reference, we provide in Table~\ref{tab:results}
the upper limits based on a 20\% signal acceptance uncertainty.

\begin{figure}[hbtp]
\begin{center}
\includegraphics[width=\cmsFigWidth]{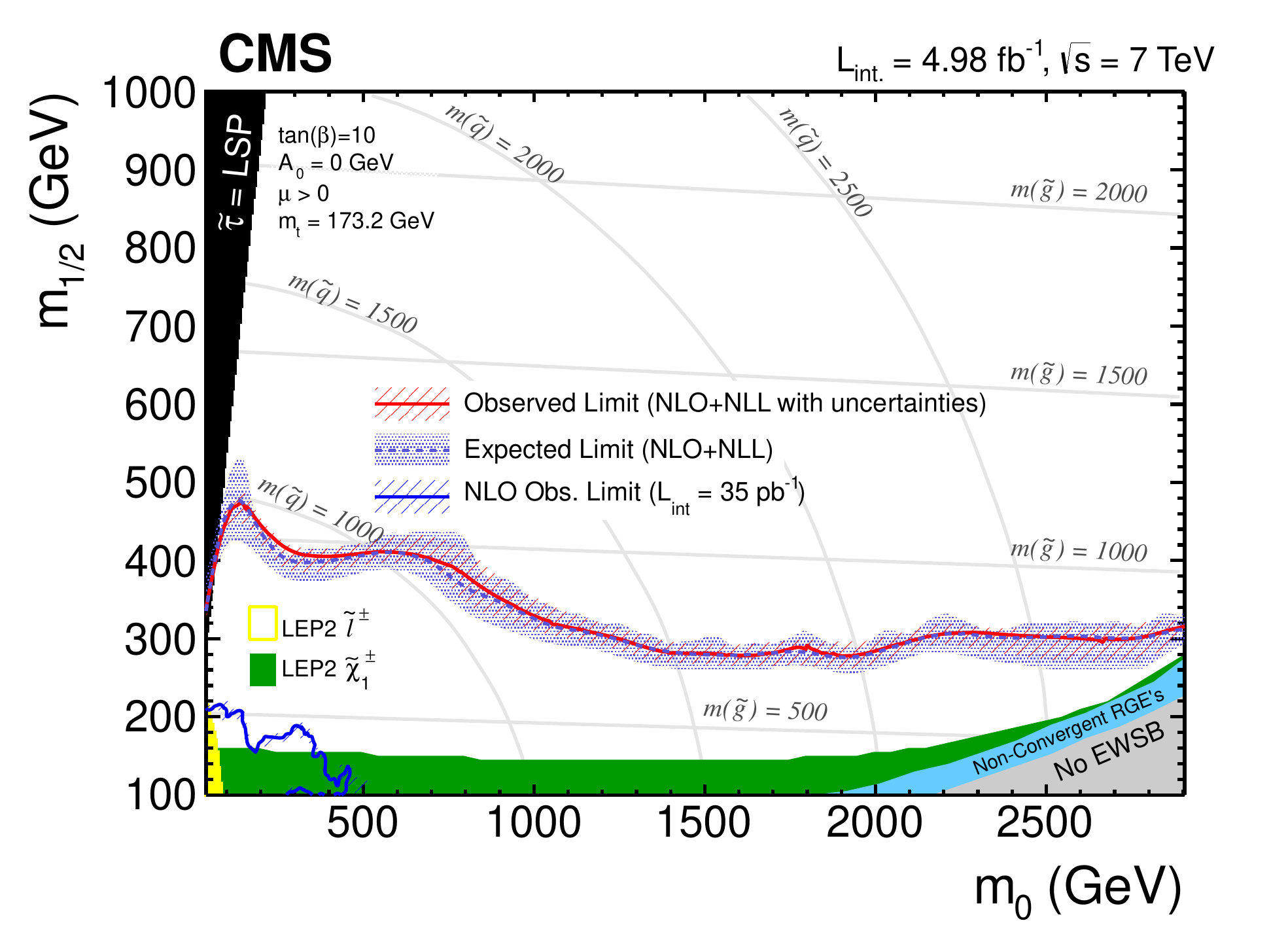}
\caption{\label{fig:cmssm}
Exclusion region, below the red curve, in the CMSSM corresponds to the observed upper limits on the number of events
from NP. The central observed curve, which includes experimental uncertainties, is obtained using high $\pt$ leptons
with $\HT > 450\gev$ and $\met> 120\gev$. The hatched region corresponds to the theoretical uncertainties on the
cross section, whereas the shaded region shows the experimental errors with $\pm 1\sigma$ variation. We also show the result
of the previous analysis~\cite{sus10004} to illustrate the improvement.}
\end{center}
\end{figure}

In order to compare our signal sensitivity to that of other searches for SUSY,
we interpret the results in the context of the CMSSM model.
We compare the observed upper limits on the number of signal events reported
in Table~\ref{tab:results} to the expected number of events in each signal region
in the CMSSM model in a plane of $(m_0,m_{1/2})$ for $\tan\beta=10$, $A_0=0$, and $\mu>0$.
For each point in the CMSSM, we choose the signal region providing the best expected upper limit on the cross section to evaluate the observed
limit; in all cases the best limit is achieved in Region~4, where
high $\pt$ leptons, large $\HT > 450\gev$, and $\met > 120\gev$ are required.
We interpret all points having mean expected values above the corresponding
observed upper limit as excluded at the 95\% CL.  For this exercise
the systematic uncertainty on the signal acceptance is re-evaluated for each point
in order for the upper limit to reflect the varying influences of the jet energy scale uncertainty.
We display the observed exclusion region in Fig.~\ref{fig:cmssm}.
For $m_0 > 1.3\TeV$,
the exclusion curve flattens out at about $m_{1/2}\sim 290$\gev, which corresponds
to a wino-like $\widetilde{\chi}^{\pm}_{1}$ mass of ${\sim}200\gev$.
The new result extends the excluded CMSSM region to gluino masses of 710\gev. This exclusion includes a -1$\sigma_{\mathrm th}$ reduction to
account for theory uncertainty~\cite{Kramer:2012bx,prospino,Kulesza:2008jb,Kulesza:2009kq,Beenakker:2009ha,Beenakker:2011fu,Beenakker:1997ut,Beenakker:2010nq,Beenakker:1999xh}
on the cross section; the limit is independent of the squark masses.

One of the challenges of signature-based searches is to convey information
in a form that can be used to
test a variety of NP models.
In Ref.~\cite{sus10004}, additional information is presented that can be used to confront
NP models in an approximate way through
generator-level simulation studies.
The approximate model of lepton, jet, and \met\ selection efficiencies
in terms of the generator-level quantities are shown to be sufficiently accurate
to reproduce the constraints on NP models that otherwise would require the
full CMS detector simulation.
The efficiency dependence can be parameterized as a function of \pt\ (expressed in \gev) as
$0.72 \{\erf [(\pt - 10)/22.5] \} + 0.22 \{ 1 - {\rm erf}[(\pt - 10)/22.5] \}$ for electrons,
$0.79 \{\erf [(\pt - 5)/19.5] \} + 0.41 \{ 1 - {\rm erf}[(\pt - 5)/19.5]\}$ for muons,
 and $0.34 {1-\exp[-0.052(\pt-15)]}$ for taus, where $\erf$ is the error function.
We studied the efficiency for an event to pass a given reconstructed \met~($\HT$) threshold
as a function of the generator-level \met~($\HT$), where in the latter case \met~is computed using neutrinos
and the LSPs and $\HT$ is the scalar sum of the transverse momenta of the partons
that satisfy the same jet selection criteria used in this analysis.
The dependences are parameterized by
$0.5 \epsilon_{\infty} \{\erf[(x - x_{1/2})/\sigma] + 1\}$,
where $x$ corresponds to the generator-level \met\ or \HT,
$\epsilon_{\infty}$ is the selection efficiency plateau at high values of $x$,
$x_{1/2}$ is the value of $x$ corresponding to half the plateau efficiency,
and $\sigma$ determines how fast the efficiency changes with $x$.
For the \HT\ selections of 200 and 450\gev,
the values of $(\epsilon_{\infty}, x_{1/2}, \sigma)$
are $(0.997, 185\gev, 99\gev)$, and $(0.992, 441\gev, 120\gev)$,
respectively.
For the \met\ selections of 50 and 120\gev, the parameters are
$(0.999, 43\gev, 39\gev)$,
and
$(0.999, 123\gev, 37\gev)$, respectively.
We tested the parameterized efficiency model in the CMSSM, and the results obtained
agree at the 15\% level with the full simulation results.

In summary, we conducted a search for physics beyond the standard model based on same-sign dileptons in the $\Pe\Pe$,
$\Pgm\Pgm$, $\Pe\Pgm$, $\Pe\Pgt$, $\Pgm\Pgt$, and $\Pgt\Pgt$ final states,
and find no evidence for an excess over the expected standard model background.
We set 95\% CL upper limits on contributions from new physics processes
based on an integrated luminosity of  \intLumi in the range of 6.2 to 16.9 events, depending on the signal search region.
These are the most restrictive limits in this particular final state to date.
We have also shown the excluded region in the CMSSM parameter space.

We wish to congratulate our colleagues in the CERN accelerator departments
for the excellent performance of the LHC machine.
We thank the technical and administrative staff at CERN and other CMS institutes,
and acknowledge support from:
FMSR (Austria); FNRS and FWO (Belgium);
CNPq, CAPES, FAPERJ, and FAPESP (Brazil);
MES (Bulgaria); CERN; CAS, MoST, and NSFC (China); COLCIENCIAS (Colombia);
MSES (Croatia); RPF (Cyprus); Academy of Sciences and NICPB (Estonia);
Academy of Finland, MEC, and HIP (Finland); CEA and CNRS/IN2P3 (France);
BMBF, DFG, and HGF (Germany); GSRT (Greece); OTKA and NKTH (Hungary);
DAE and DST (India); IPM (Iran); SFI (Ireland); INFN (Italy); NRF and WCU (Korea);
LAS (Lithuania); CINVESTAV, CONACYT, SEP, and UASLP-FAI (Mexico); MSI (New Zealand);
PAEC (Pakistan); SCSR (Poland); FCT (Portugal);
JINR (Armenia, Belarus, Georgia, Ukraine, Uzbekistan);
MST, MAE and RFBR (Russia); MSTD (Serbia); MICINN and CPAN (Spain);
Swiss Funding Agencies (Switzerland); NSC (Taipei); TUBITAK and TAEK (Turkey);
STFC (United Kingdom); DOE and NSF (USA).

\bibliography{auto_generated}   

\providecommand{\href}[2]{#2}\begingroup\raggedright\begin{thebibliography}{10}%
\makeatletter
\providecommand{\hrefCMSnoop }[0]{\@secondoftwo}%
\makeatother
\providecommand{\doi}{\texttt{doi:}\begingroup \urlstyle{tt}\Url}

\bibitem{Fayet:1976cr}
\hrefCMSnoop {} {P.~Fayet and S.~Ferrara, ``{Supersymmetry}'',} \textit{ Phys.
  Rept.} \textbf{ 32} (1977) 249,
  \href{http://dx.doi.org/10.1016/0370-1573(77)90066-7}{\doi{10.1016/0370-1573(77)90066-7}},
\href{http://www.arXiv.org/abs/1103.0953}{\texttt{ arXiv:1103.0953}}.

\bibitem{AtlasHad}
\hrefCMSnoop {} {{ ATLAS} Collaboration, ``{Search for new phenomena in final
  states with large jet multiplicities and missing transverse momentum using
  $\sqrt{s}$ = 7 TeV pp collisions with the ATLAS detector}'',} \textit{ JHEP}
  \textbf{ 11} (2011) 099,
  \href{http://dx.doi.org/10.1007/JHEP11(2011)099}{\doi{10.1007/JHEP11(2011)099}},
\href{http://www.arXiv.org/abs/1110.2299}{\texttt{ arXiv:1110.2299}}.

\bibitem{CMSHad}
\hrefCMSnoop {} {{ CMS} Collaboration, ``{Search for Supersymmetry at the LHC
  in Events with Jets and Missing Transverse Energy}'',} \textit{ Phys. Rev.
  Lett.} \textbf{ 107} (2011) 221804,
  \href{http://dx.doi.org/10.1103/PhysRevLett.107.221804}{\doi{10.1103/PhysRevLett.107.221804}},
\href{http://www.arXiv.org/abs/1109.2352}{\texttt{ arXiv:1109.2352}}.

\bibitem{Atlas1lep}
\hrefCMSnoop {} {{ ATLAS} Collaboration, ``{Search for supersymmetry in final
  states with jets, missing transverse momentum and one isolated lepton in
  $\sqrt{s}$ = 7 TeV pp collisions using 1 fb$^{-1}$ of ATLAS data}'',}
  \textit{ Phys. Rev. D} \textbf{ 85} (2012) 012006,
  \href{http://dx.doi.org/10.1103/PhysRevD.85.012006}{\doi{10.1103/PhysRevD.85.012006}},
\href{http://www.arXiv.org/abs/1109.6606}{\texttt{ arXiv:1109.6606}}.

\bibitem{CMS1lep}
\hrefCMSnoop {} {{ CMS} Collaboration, ``{Search for supersymmetry in pp
  collisions at $\sqrt{s}$ = 7 TeV in events with a single lepton, jets, and
  missing transverse momentum}'',} \textit{ JHEP} \textbf{ 08} (2011) 156,
  \href{http://dx.doi.org/10.1007/JHEP08(2011)156}{\doi{10.1007/JHEP08(2011)156}},
\href{http://www.arXiv.org/abs/1107.1870}{\texttt{ arXiv:1107.1870}}.

\bibitem{AtlasOS}
\hrefCMSnoop {} {{ ATLAS} Collaboration, ``{Searches for supersymmetry with the
  ATLAS detector using final states with two leptons and missing transverse
  momentum in $\sqrt{s}$ = 7 TeV proton-proton collisions}'',} \textit{ Phys.
  Lett. B} \textbf{ 709} (2012) 137,
  \href{http://dx.doi.org/10.1016/j.physletb.2012.01.076}{\doi{10.1016/j.physletb.2012.01.076}},
\href{http://www.arXiv.org/abs/1110.6189}{\texttt{ arXiv:1110.6189}}.

\bibitem{CMSOS}
\hrefCMSnoop {} {{ CMS} Collaboration, ``{Search for physics beyond the
  standard model in opposite-sign dilepton events at $\sqrt{s} = 7$ TeV}'',}
  \textit{ JHEP} \textbf{ 06} (2011) 026,
  \href{http://dx.doi.org/10.1007/JHEP06(2011)026}{\doi{10.1007/JHEP06(2011)026}},
\href{http://www.arXiv.org/abs/1103.1348}{\texttt{ arXiv:1103.1348}}.

\bibitem{sus10004}
\hrefCMSnoop {} {{ CMS} Collaboration, ``{Search for new physics with same-sign
  isolated dilepton events with jets and missing transverse energy at the
  LHC}'',} \textit{ JHEP} \textbf{ 06} (2011) 077,
  \href{http://dx.doi.org/10.1007/JHEP06(2011)077}{\doi{10.1007/JHEP06(2011)077}},
  \href{http://www.arXiv.org/abs/1104.3168}{\texttt{ arXiv:1104.3168}}.

\bibitem{CMSmult}
\hrefCMSnoop {} {{ CMS} Collaboration, ``{Search for physics beyond the
  standard model using multilepton signatures in pp collisions at $\sqrt{s}$ =
  7 TeV}'',} \textit{ Phys. Lett. B} \textbf{ 704} (2011) 411,
  \href{http://dx.doi.org/10.1016/j.physletb.2011.09.047}{\doi{10.1016/j.physletb.2011.09.047}},
\href{http://www.arXiv.org/abs/1106.0933}{\texttt{ arXiv:1106.0933}}.

\bibitem{Atlasgamma}
\hrefCMSnoop {} {{ ATLAS} Collaboration, ``{Search for diphoton events with
  large missing transverse momentum in 1 fb$^{-1}$ of 7 TeV proton-proton
  collision data with the ATLAS Detector}'',} \textit{ Phys. Lett. B} \textbf{
  710} (2012) 519,
  \href{http://dx.doi.org/10.1016/j.physletb.2012.02.054}{\doi{10.1016/j.physletb.2012.02.054}},
\href{http://www.arXiv.org/abs/1111.4116}{\texttt{ arXiv:1111.4116}}.

\bibitem{CMSgamma}
\hrefCMSnoop {} {{ CMS} Collaboration, ``{Search for Supersymmetry in $pp$
  Collisions at $\sqrt{s}$ = 7 TeV in Events with Two Photons and Missing
  Transverse Energy}'',} \textit{ Phys. Rev. Lett.} \textbf{ 106} (2011)
  211802,
  \href{http://dx.doi.org/10.1103/PhysRevLett.106.211802}{\doi{10.1103/PhysRevLett.106.211802}},
\href{http://www.arXiv.org/abs/1103.0953}{\texttt{ arXiv:1103.0953}}.

\bibitem{Barnett:1993ea}
\hrefCMSnoop {} {R.~M. Barnett, J.~F. Gunion, and H.~E. Haber, ``Discovering
  supersymmetry with like sign dileptons'',} \textit{ Phys. Lett. B} \textbf{
  315} (1993) 349,
  \href{http://dx.doi.org/10.1016/0370-2693(93)91623-U}{\doi{10.1016/0370-2693(93)91623-U}},
  \href{http://www.arXiv.org/abs/9306204}{\texttt{ arXiv:9306204}}.

\bibitem{Guchait:1994zk}
\hrefCMSnoop {} {M.~Guchait and D.~P. Roy, ``Like sign dilepton signature for
  gluino production at the CERN LHC including top quark and Higgs boson
  effects'',} \textit{ Phys. Rev. D} \textbf{ 52} (1995) 133,
  \href{http://dx.doi.org/10.1103/PhysRevD.52.133}{\doi{10.1103/PhysRevD.52.133}},
  \href{http://www.arXiv.org/abs/9412329}{\texttt{ arXiv:9412329}}.

\bibitem{Baer:1995va}
H.~Baer\hrefCMSnoop {} { {et~al.}, ``Signals for minimal supergravity at the
  CERN Large Hadron Collider II: Multilepton channels'',} \textit{ Phys. Rev.
  D} \textbf{ 53} (1996) 6241,
  \href{http://dx.doi.org/10.1103/PhysRevD.53.6241}{\doi{10.1103/PhysRevD.53.6241}},
  \href{http://www.arXiv.org/abs/9512383}{\texttt{ arXiv:9512383}}.

\bibitem{Cheng:2002ab}
\hrefCMSnoop {} {H.~C. Cheng, K.~T. Matchev, and M.~Schmaltz, ``Bosonic
  supersymmetry? Getting fooled at the CERN LHC'',} \textit{ Phys. Rev. D}
  \textbf{ 66} (2002) 056006,
  \href{http://dx.doi.org/10.1103/PhysRevD.66.056006}{\doi{10.1103/PhysRevD.66.056006}},
  \href{http://www.arXiv.org/abs/0205314}{\texttt{ arXiv:0205314}}.

\bibitem{Jung:2010}
\hrefCMSnoop {} {S.~Jung, H.~Murayama, and A.~Pierce, ``{Top quark
  forward-backward asymmetry from new t-channel physics}'',} \textit{ Phys.
  Rev. D} \textbf{ 81} (2010) 015004,
  \href{http://dx.doi.org/10.1103/PhysRevD.81.015004}{\doi{10.1103/PhysRevD.81.015004}},
\href{http://www.arXiv.org/abs/0907.4112}{\texttt{ arXiv:0907.4112}}.

\bibitem{Berger:2011}
E.~L. Berger\hrefCMSnoop {} { {et~al.}, ``{Top Quark Forward-Backward Asymmetry
  and Same-Sign Top Quark Pairs}'',} \textit{ Phys. Rev. Lett.} \textbf{ 106}
  (2011) 201801,
  \href{http://dx.doi.org/10.1103/PhysRevLett.106.201801}{\doi{10.1103/PhysRevLett.106.201801}},
\href{http://www.arXiv.org/abs/1101.5625}{\texttt{ arXiv:1101.5625}}.

\bibitem{Contino:2008hi}
\hrefCMSnoop {} {R.~Contino and G.~Servant, ``Discovering the top partners at
  the LHC using same-sign dilepton final states'',} \textit{ JHEP} \textbf{ 06}
  (2008) 026,
  \href{http://dx.doi.org/10.1088/1126-6708/2008/06/026}{\doi{10.1088/1126-6708/2008/06/026}},
  \href{http://www.arXiv.org/abs/0801.1679}{\texttt{ arXiv:0801.1679}}.

\bibitem{Keung:1983uu}
\hrefCMSnoop {} {W.-Y. Keung and G.~Senjanovic, ``{Majorana neutrinos and the
  production of the right-handed charged gauge boson.}'',} \textit{ Phys. Rev.
  Lett.} \textbf{ 50} (1983) 1427,
\href{http://dx.doi.org/10.1103/PhysRevLett.50.1427}{\doi{10.1103/PhysRevLett.50.1427}}.

\bibitem{Han:2009}
\hrefCMSnoop {} {Y.~Bai and Z.~Han, ``Top-antitop and Top-top Resonances in the
  Dilepton Channel at the CERN LHC'',} \textit{ JHEP} \textbf{ 04} (2009) 056,
  \href{http://dx.doi.org/10.1088/1126-6708/2009/04/056}{\doi{10.1088/1126-6708/2009/04/056}},
  \href{http://www.arXiv.org/abs/0809.4487}{\texttt{ arXiv:0809.4487}}.

\bibitem{CMS:2008zzk}
\hrefCMSnoop {} {{ CMS} Collaboration, ``{The CMS experiment at the CERN
  LHC}'',} \textit{ JINST} \textbf{ 3} (2008) S08004,
\href{http://dx.doi.org/10.1088/1748-0221/3/08/S08004}{\doi{10.1088/1748-0221/3/08/S08004}}.

\bibitem{cmssm}
G.~L. Kane\hrefCMSnoop {} { {et~al.}, ``Study of constrained minimal
  supersymmetry'',} \textit{ Phys. Rev. D} \textbf{ 49} (1994) 6173,
  \href{http://dx.doi.org/10.1103/PhysRevD.49.6173}{\doi{10.1103/PhysRevD.49.6173}},
  \href{http://www.arXiv.org/abs/9312272}{\texttt{ arXiv:9312272}}.

\bibitem{MUOPAS}
\href {http://cdsweb.cern.ch/record/1279140} {{ CMS} Collaboration,
  ``Performance of muon identification in pp collisions at $\sqrt{s}$ = 7
  {TeV}'',} CMS Physics Analysis Summary CMS-PAS-MUO-10-002, (2010).

\bibitem{EGMPAS}
\href {http://cdsweb.cern.ch/record/1299116} {{ CMS} Collaboration, ``Electron
  Reconstruction and Identification at $\sqrt{s} = 7$ {TeV}'',} CMS Physics
  Analysis Summary CMS-PAS-EGM-10-004, (2010).

\bibitem{PFT-10-XXX}
\hrefCMSnoop {} {{ CMS} Collaboration, ``{Performance of $\tau$ lepton
  reconstruction and identification in CMS}'',} \textit{ JINST} \textbf{ 7}
  (2012) P01001,
  \href{http://dx.doi.org/10.1088/1748-0221/7/01/P01001}{\doi{10.1088/1748-0221/7/01/P01001}},
\href{http://www.arXiv.org/abs/1109.6034}{\texttt{ arXiv:1109.6034}}.

\bibitem{PFT-09-001}
\href {http://cdsweb.cern.ch/record/1194487} {{ CMS} Collaboration,
  ``Particle--Flow Event Reconstruction in {CMS} and Performance for Jets,
  Taus, and {\MET}'',} CMS Physics Analysis Summary CMS-PAS-PFT-09-001, (2009).

\bibitem{PFT-10-002}
\href {http://cdsweb.cern.ch/record/1279341} {{ CMS} Collaboration,
  ``Commissioning of the Particle-Flow Reconstruction in Minimum-Bias and Jet
  Events from {\Pp\Pp} Collisions at 7 {TeV}'',} CMS Physics Analysis Summary
  CMS-PAS-PFT-10-002, (2010).

\bibitem{anti-kt}
\hrefCMSnoop {} {M.~Cacciari, G.~P. Salam, and G.~Soyez, ``The anti-$k_T$ jet
  clustering algorithm'',} \textit{ JHEP} \textbf{ 04} (2008) 063,
  \href{http://dx.doi.org/10.1088/1126-6708/2008/04/063}{\doi{10.1088/1126-6708/2008/04/063}},
  \href{http://www.arXiv.org/abs/0802.1189}{\texttt{ arXiv:0802.1189}}.

\bibitem{MadGraph}
J.~Alwall\hrefCMSnoop {} { {et~al.}, ``MadGraph/MadEvent v4: the new web
  generation'',} \textit{ JHEP} \textbf{ 09} (2007) 028,
  \href{http://dx.doi.org/10.1088/1126-6708/2007/09/028}{\doi{10.1088/1126-6708/2007/09/028}},
  \href{http://www.arXiv.org/abs/0706.2334}{\texttt{ arXiv:0706.2334}}.

\bibitem{PYTHIA}
\hrefCMSnoop {} {T.~Sj{\"o}strand, S.~Mrenna, and P.~Skands, ``PYTHIA 6.4
  physics and manual'',} \textit{ JHEP} \textbf{ 05} (2006) 026,
  \href{http://dx.doi.org/10.1088/1126-6708/2006/05/026}{\doi{10.1088/1126-6708/2006/05/026}},
  \href{http://www.arXiv.org/abs/0603175}{\texttt{ arXiv:0603175}}.

\bibitem{PTDR2}
\hrefCMSnoop {} {{CMS Collaboration}, ``{CMS} technical design report, volume
  {II}: {Physics} performance'',} \textit{ J. Phys. G} \textbf{ 34} (2007) 995,
\href{http://dx.doi.org/10.1088/0954-3899/34/6/S01}{\doi{10.1088/0954-3899/34/6/S01}}.

\bibitem{lumi}
\href {http://cdsweb.cern.ch/record/1434360} {{ CMS} Collaboration, ``Absolute
  Calibration of the Luminosity Measurement at {CMS}: {W}inter 2012 Update'',}
  CMS Physics Analysis Summary CMS-PAS-SMP-12-008, (2012).

\bibitem{Read:2002hq}
\hrefCMSnoop {} {A.~L. Read, ``{Presentation of search results: The CL(s)
  technique}'',} \textit{ J. Phys. G} \textbf{ 28} (2002) 2693,
\href{http://dx.doi.org/10.1088/0954-3899/28/10/313}{\doi{10.1088/0954-3899/28/10/313}}.

\bibitem{Junk:1999kv}
\hrefCMSnoop {} {T.~Junk, ``{Confidence level computation for combining
  searches with small statistics}'',} \textit{ Nucl. Instrum. Meth. A} \textbf{
  434} (1999) 435,
  \href{http://dx.doi.org/10.1016/S0168-9002(99)00498-2}{\doi{10.1016/S0168-9002(99)00498-2}},
\href{http://www.arXiv.org/abs/9902006}{\texttt{ arXiv:9902006}}.

\bibitem{ATL-PHYS-PUB-2011-011}
\href {http://cdsweb.cern.ch/record/1379837} {{ATLAS and CMS Collaborations,
  LHC Higgs Combination Group}, ``Procedure for the {LHC} {H}iggs boson search
  combination in {S}ummer 2011'',} ATL-PHYS-PUB 2011-11, (2011).
\newblock {CMS NOTE 2011-005}.

\bibitem{Kramer:2012bx}
M.~Kramer\hrefCMSnoop {} { {et~al.}, ``{Supersymmetry production cross sections
  in pp collisions at sqrt{s} = 7 TeV}'',} (2012).
\href{http://www.arXiv.org/abs/1206.2892}{\texttt{ arXiv:1206.2892}}.

\bibitem{prospino}
W.~Beenakker\hrefCMSnoop {} { {et~al.}, ``Squark and Gluino Production at
  Hadron Colliders'',} \textit{ Nucl. Phys.} \textbf{ B 492} (1997) 51,
  \href{http://dx.doi.org/10.1016/S0550-3213(97)80027-2}{\doi{10.1016/S0550-3213(97)80027-2}}.

\bibitem{Kulesza:2008jb}
\hrefCMSnoop {} {A.~Kulesza and L.~Motyka, ``{Threshold resummation for
  squark-antisquark and gluino-pair production at the LHC}'',} \textit{ Phys.
  Rev. Lett.} \textbf{ 102} (2009) 111802,
  \href{http://dx.doi.org/10.1103/PhysRevLett.102.111802}{\doi{10.1103/PhysRevLett.102.111802}},
\href{http://www.arXiv.org/abs/0807.2405}{\texttt{ arXiv:0807.2405}}.

\bibitem{Kulesza:2009kq}
\hrefCMSnoop {} {A.~Kulesza and L.~Motyka, ``{Soft gluon resummation for the
  production of gluino-gluino and squark-antisquark pairs at the LHC}'',}
  \textit{ Phys. Rev. D} \textbf{ 80} (2009) 095004,
  \href{http://dx.doi.org/10.1103/PhysRevD.80.095004}{\doi{10.1103/PhysRevD.80.095004}},
\href{http://www.arXiv.org/abs/0905.4749}{\texttt{ arXiv:0905.4749}}.

\bibitem{Beenakker:2009ha}
W.~Beenakker\hrefCMSnoop {} { {et~al.}, ``{Soft-gluon resummation for squark
  and gluino hadroproduction}'',} \textit{ JHEP} \textbf{ 0912} (2009) 041,
  \href{http://dx.doi.org/10.1088/1126-6708/2009/12/041}{\doi{10.1088/1126-6708/2009/12/041}},
\href{http://www.arXiv.org/abs/0909.4418}{\texttt{ arXiv:0909.4418}}.

\bibitem{Beenakker:2011fu}
W.~Beenakker\hrefCMSnoop {} { {et~al.}, ``{Squark and gluino
  hadroproduction}'',} \textit{ Int. J. Mod. Phys. A} \textbf{ 26} (2011) 2637,
  \href{http://dx.doi.org/10.1142/S0217751X11053560}{\doi{10.1142/S0217751X11053560}},
\href{http://www.arXiv.org/abs/1105.1110}{\texttt{ arXiv:1105.1110}}.

\bibitem{Beenakker:1997ut}
W.~Beenakker\hrefCMSnoop {} { {et~al.}, ``{Stop production at hadron
  colliders}'',} \textit{ Nucl. Phys. B} \textbf{ 515} (1998) 3,
  \href{http://dx.doi.org/10.1016/S0550-3213(98)00014-5}{\doi{10.1016/S0550-3213(98)00014-5}},
\href{http://www.arXiv.org/abs/hep-ph/9710451}{\texttt{ arXiv:hep-ph/9710451}}.

\bibitem{Beenakker:2010nq}
W.~Beenakker\hrefCMSnoop {} { {et~al.}, ``{Supersymmetric top and bottom squark
  production at hadron colliders}'',} \textit{ JHEP} \textbf{ 1008} (2010) 098,
  \href{http://dx.doi.org/10.1007/JHEP08(2010)098}{\doi{10.1007/JHEP08(2010)098}},
\href{http://www.arXiv.org/abs/1006.4771}{\texttt{ arXiv:1006.4771}}.

\bibitem{Beenakker:1999xh}
W.~Beenakker\hrefCMSnoop {} { {et~al.}, ``{The Production of charginos /
  neutralinos and sleptons at hadron colliders}'',} \textit{ Phys. Rev. Lett.}
  \textbf{ 83} (1999) 3780,
  \href{http://dx.doi.org/10.1103/PhysRevLett.83.3780,
  10.1103/PhysRevLett.100.029901}{\doi{10.1103/PhysRevLett.83.3780,
  10.1103/PhysRevLett.100.029901}},
\href{http://www.arXiv.org/abs/hep-ph/9906298}{\texttt{ arXiv:hep-ph/9906298}}.

\end{thebibliography}\endgroup

\cleardoublepage \appendix\section{The CMS Collaboration \label{app:collab}}\begin{sloppypar}\hyphenpenalty=5000\widowpenalty=500\clubpenalty=5000\textbf{Yerevan Physics Institute,  Yerevan,  Armenia}\\*[0pt]
S.~Chatrchyan, V.~Khachatryan, A.M.~Sirunyan, A.~Tumasyan
\vskip\cmsinstskip
\textbf{Institut f\"{u}r Hochenergiephysik der OeAW,  Wien,  Austria}\\*[0pt]
W.~Adam, T.~Bergauer, M.~Dragicevic, J.~Er\"{o}, C.~Fabjan\cmsAuthorMark{1}, M.~Friedl, R.~Fr\"{u}hwirth\cmsAuthorMark{1}, V.M.~Ghete, J.~Hammer, N.~H\"{o}rmann, J.~Hrubec, M.~Jeitler\cmsAuthorMark{1}, W.~Kiesenhofer, V.~Kn\"{u}nz, M.~Krammer\cmsAuthorMark{1}, D.~Liko, I.~Mikulec, M.~Pernicka$^{\textrm{\dag}}$, B.~Rahbaran, C.~Rohringer, H.~Rohringer, R.~Sch\"{o}fbeck, J.~Strauss, A.~Taurok, P.~Wagner, W.~Waltenberger, G.~Walzel, E.~Widl, C.-E.~Wulz\cmsAuthorMark{1}
% ...
\vskip\cmsinstskip
\dag:~Deceased\\
1:~~Also at Vienna University of Technology, Vienna, Austria\\
% ...

\end{sloppypar}
\end{document}